\documentclass[aps,prb,reprint,showpacs,superscriptaddress]{revtex4-1}
\usepackage[english]{babel}
\usepackage{amsmath,amssymb,bbm,graphicx,color,comment,txfonts}
\usepackage[bookmarks=true,colorlinks,citecolor=blue,urlcolor=blue]{hyperref}
\usepackage{dsfont}
\usepackage[normalem]{ulem}
\usepackage{physics}
\begin{document}

\date{\today}

\title{Entanglement phases, localization and multifractality of monitored free fermions in two dimensions}

\author{K. Chahine}
\affiliation{Institut f\"ur Theoretische Physik, Universit\"at zu K\"oln, D-50937 Cologne, Germany}
\author{M. Buchhold}
\affiliation{Institut f\"ur Theoretische Physik, Universit\"at zu K\"oln, D-50937 Cologne, Germany}

\begin{abstract}
We investigate the entanglement structure and wave function characteristics of continuously monitored free fermions with 
U$(1)$-symmetry in two spatial dimensions (2D). By deriving the exact fermion replica-quantum master equation, we line out two approaches: (i) a nonlinear sigma model analogous to disordered free fermions, resulting in an SU$(R)$-symmetric field theory of symmetry class AIII in (2+1) space-time dimensions, or (ii) for bipartite lattices, third quantization leading to a non-Hermitian SU$(2R)$-symmetric Hubbard model. Using exact numerical simulations, we explore the phenomenology of the entanglement transition in 2D monitored fermions, examining entanglement entropy and wave function inverse participation ratio. At weak monitoring, we observe characteristic $L\log L$ entanglement growth and multifractal dimension $D_q=2$, resembling a metallic Fermi liquid. Under strong monitoring, wave functions localize and the entanglement saturates towards an area law. Between these regimes, we identify a high-symmetry point exhibiting both entanglement growth indicative of emergent conformal invariance and maximal multifractal behavior. 
While this multifractal behavior aligns with the nonlinear sigma model of the Anderson transition, the emergent conformal invariance is an unexpected feature not typically associated with Anderson localization. These discoveries add a new dimension to the study of 2D monitored fermions and underscore the need to further explore the connection between non-unitary quantum dynamics in $D$ dimensions and quantum statistical mechanics in $D+1$ dimensions.
\end{abstract}

\maketitle
\section{Introduction} The advances in realizing quantum devices with unitary evolution and mid-circuit measurements have put focus on a novel type of quantum dynamics: the competition between scrambling and localization of quantum information, caused by the non-commutivity of delocalizing unitary dynamics and localizing non-unitary measurements. The genuine quantum mechanical competition culminates in a monitoring-induced phase transition (MIPT) in the entanglement entropy. Two types of protocols have crystallized: \emph{quantum circuits} utilizing discrete unitary gates and projective  measurements~\cite{
Skinner2019,Fisher2018,Li2019b,gullans2019,choi2020prl,Jian2020,fan2020selforganized,lifisher2021,nahum2021prxq, jian2021syk,  bao2021symmetry,turkeshi2021measurementinduced, Ivanov2020,Buonaiuto2021,  Zabalo2020,ippoliti2021,circuitreview,Romito2020, Schomerus2019, turkeshi2022measurement,Popperl,alba2022,KhemaniKitaev,Zhu2022,Zhu2023,ippoliti2020,Klocke2022,Klocke2023,SagarKitaev,Morral2022} and \emph{monitored Hamiltonians}, augmenting a continuous unitary evolution by measurements~\cite{Cao2019,alberton2021enttrans,Bernard_2018,buchhold2021effective,poboiko,Ladewig2022,minato2021fate,szyniszewski2023disordered, kells2023,doggen2021generalized, tsitsishvili2023measurement, turkeshi2022entanglement,piccitto2022, Mueller2022,oshima2023,paviglianiti2023multipartite, minoguchi2021continuous, tirrito2023, xing2023interactions, turkeshi2023entanglement,fuji2020,Altland2022, turkeshi2023density,Yang2023Keldysh,Coppola}. 

Recently, MIPTs in $D$ dimensions have been linked to localization-delocalization transitions in $D+1$-dimensional disordered quantum systems~\cite{Zabalo2022,sierant2022,Klocke2023,Bao_2020,buchhold2021effective,poboiko, szyniszewski2023disordered,boorman2022, Behrends2022, yamamoto2023,JianNLS,Fava2023}, comparing random, non-unitary measurements in a $D$-dimensional evolution with static disorder  in $D + 1$ space-time dimensions. However, the primary focus has been on dynamics in one spatial dimension, where an exponentially diverging correlation length of fermions with U$(1)$-symmetry complicates the clear identification of the MIPT~\cite{Cao2019,alberton2021enttrans,buchhold2021effective,szyniszewski2023disordered,minato2021fate}. 

\begin{figure}[th!]
\includegraphics[width=\linewidth]{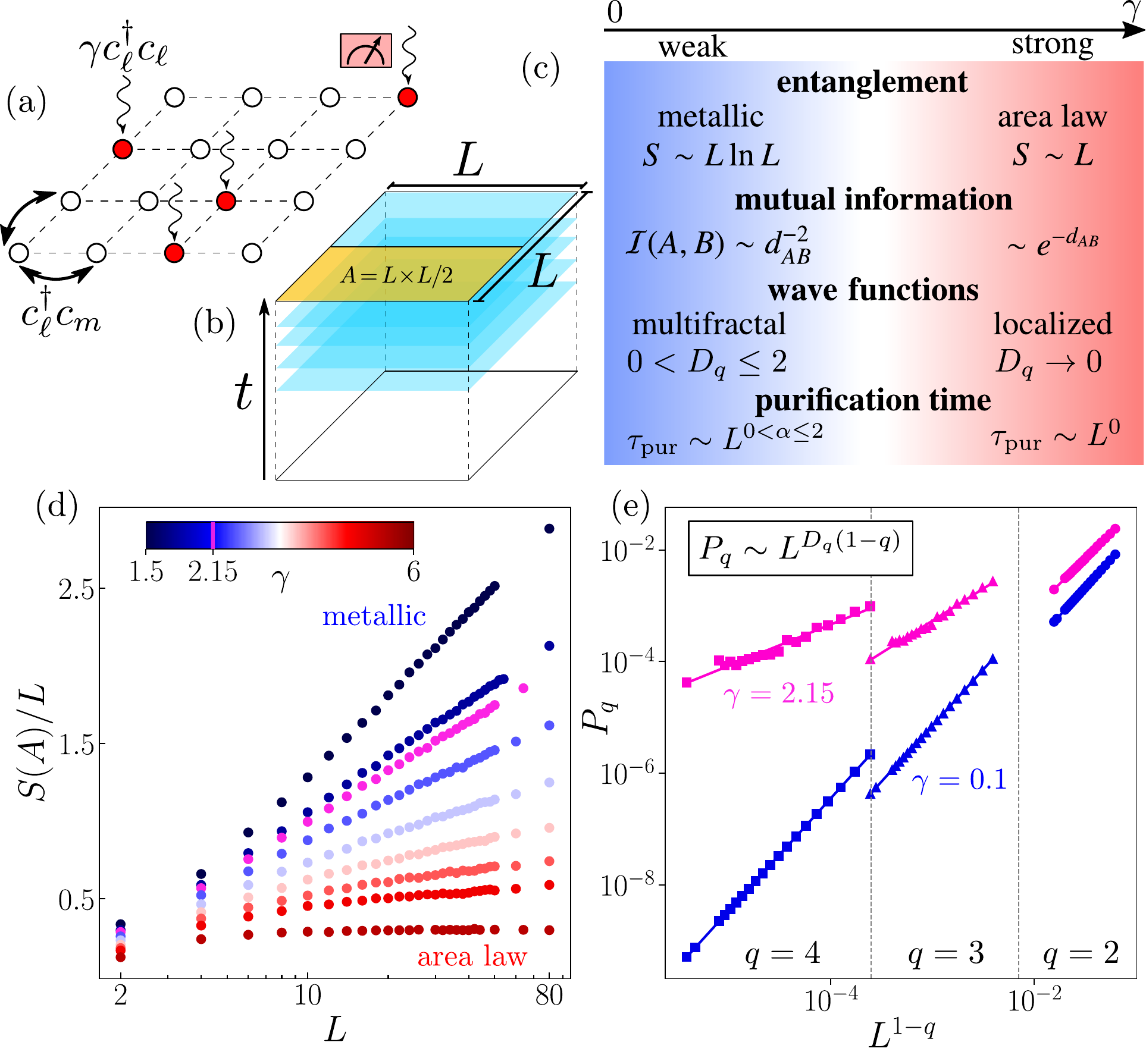}
\caption{\label{fig:overview} {\bf Measurement-induced phase transition of monitored fermions in 2D}. (a) Free fermions hopping on an $L\times L$ square lattice are continuously monitored with rate $\gamma$. (b) Sketch of the time evolution. Blue sheets represent snapshot wave functions from which observables, including the half-system entanglement entropy $S(A)$ for strips $A=L\times L/2$ (yellow) are obtained (c) Schematic phase diagram showing observables and their characteristic behavior. (d) At weak monitoring $S(A)\sim L\log L$, while saturating to an area law $\sim L$ for strong monitoring. (d) Multifractal exponent $D_q$ of the inverse participation ratio $P_q$: Wave function  multifractality is observed around $\gamma=2.15$ (pink, $D_2=1.80(9)$, $D_3=1.1(5)$ and $D_4=0.7(4)$) and metallic behavior for weak monitoring (blue, $D_q= 2$ for all $q$).}
\end{figure}

Here we explore this link in a new arena with a rich phenomenology: we study continuously monitored, U$(1)$-symmetric free fermions in $D=2$ spatial dimensions, as illustrated in Fig.~\ref{fig:overview}a,b. We use a two-pronged approach, combining analytical tools and large scale numerical simulations. An intriguing picture is provided by fermion Keldysh replica field theory: a direct mapping of the $R$-replicated lattice to a continuum theory yields a description reminiscent of free disordered fermions with a dissipative but \emph{space-time local} four-fermion vertex~\cite{Kamenev2011}. Expanding around the metallic saddle point, we derive a nonlinearinear sigma model (NL$\sigma$M) with SU$(R)$ symmetry for the Goldstone modes, as previously obtained for projective measurements~\cite{poboiko}. However, we show for bipartite lattices that space-time locality enables an alternative mapping of the replica Keldysh master equation to an SU$(2R)$-symmetric, non-Hermitian Hubbard model with a dissipative interaction vertex in third quantization. This anticipates a rich phenomenology for MIPTs in $2+1$ dimensions, which we explore with numerical simulations.

We numerically explore the phenomenology of the MIPT, a localization-delocalization transition, whose features are summarized in Fig.~\ref{fig:overview}c. At weak monitoring, the half-system entanglement entropy exhibits 
$L\log L$-growth, characteristic of a metallic phase, which transitions to a localized, area law phase at strong monitoring, see Fig.~\ref{fig:overview}d. Between these phases, we identify a unique high symmetry point exhibiting scaling consistent with a conformal field theory (CFT). This finding contrasts with Anderson localization, which predicts area-law scaling at the transition~\cite{EversMirlin}. Interestingly, at the high symmetry point, the fermion wave functions display multifractal behavior, shown in Fig.~\ref{fig:overview}e, that aligns with the nonlinear sigma model (NL$\sigma$M) predictions for class AI in three dimensions ~\cite{sierant2022,Zabalo2022,turkeshi2023measuring,niroula2023phase,Mildenberger,EversMirlin}. We show that the purification time of a mixed initial state is determined by the multifractal exponent. For free fermions entanglement arises solely from particle number fluctuations, which links the purification process to the multifractal behavior at a MIPT with simultaneous charge sharpening~\cite{Agraval2021,Shayan2022,Barratt2023}.


\section{Theoretical background}
\subsection{Microscopic model} We consider fermions on a half-filled 2D $L\times L$ square lattice, with creation and annihilation operators $\{\hat{c}_\ell^{\phantom{^\dagger}}$, $\hat{c}_{\ell'}^\dagger\}=\delta_{\ell,\ell'}$ on lattice sites $\ell, \ell'$. The fermions undergo coherent nearest-neighbour hopping with Hamiltonian $\hat H=-\sum_{\langle \ell,m\rangle}\hat{c}_m^\dagger\hat{c}_\ell^{\phantom{^\dagger}}+\hat{c}_\ell^\dagger\hat{c}_m^{\phantom{^\dagger}}$. Simultaneously, the particle number $\hat{n}_\ell$ at each site is continuously monitored with rate $\gamma$. Monitoring is implemented by the generalized projector~\cite{book_QO_Walls,book_QControl_Wiseman}
\begin{align}\label{eq:proj}
\hat M(\{J_{\ell,t}\})=\prod_\ell[\tfrac{2\gamma dt}{\pi}]^{\frac{1}{4}}\exp[-\gamma dt (J_{\ell,t}-\hat n_\ell)^2].
\end{align}
The \emph{unnormalized} wave function  $|\tilde\psi_t\rangle$ evolves at time $t$ $ |\tilde\psi_{t+dt}\rangle=\text{exp} (-i\hat H dt) \hat M_{\ell}(\{J_{\ell,t}\})|\tilde\psi_{t}\rangle$
after measurement outcoms $J_{\ell,t}\in\mathds{R}$ were recorded. The normalized wave function $|\psi_{t}\rangle= |\tilde\psi_{t}\rangle p(\{J_{\ell,t}\})^{-\frac{1}{2}}$ is obtained via the Born probability for measuring the stream $\{J_{\ell,t'<t}\}$ for all sites $\ell$ and times $t'<t$,
\begin{align}
p(\{J_{\ell,t+dt}\})=\langle \psi_{t}|\hat M_{\ell}(\{J_{\ell,t}\})^2|\psi_{t}\rangle p(\{J_{\ell,t}\})=\langle\tilde\psi_{t+dt}|\tilde\psi_{t+dt}\rangle.
\end{align} 

An equivalent formulation emerges in the limit of infinitesimal time steps $dt\to0^+$. For a single wave function, the evolution of the normalized wave function $|\psi_{t}\rangle\to |\psi_{t+dt}\rangle$ in an infinitesimal time step is composed of three steps. First, the measurement outcomes $J_{\ell,t}$ are drawn from the Born probabilities $p(\{J_{\ell,t}\})$. Second, the unnormalized wave function is computed $|\tilde\psi_{t+dt}\rangle=\text{exp} (-i\hat H dt) \hat M_{\ell}(\{J_{\ell,t}\})|\psi_{t}\rangle$ from the normalized one. Then $ |\psi_{t+dt}\rangle=|\tilde\psi_{t+dt}\rangle || |\tilde\psi_{t+dt}\rangle||^{-\frac{1}{2}}$ normalization is performed. For $dt\to 0^+$, the variables $J_{\ell,t}$ become Gaussian distributed with mean $\overline{J_{\ell,t}}=\langle\hat n_\ell\rangle_t$ and variance $\text{var}(J_{\ell,t})=\tfrac{1}{4\gamma dt}$. This is readily verified by taking the $m$-th moment of $J_{\ell,t}$,
\begin{align}
    \overline{J_{\ell,t}^m}=\int_{-\infty}^\infty dJ_{\ell,t}\ J_{\ell,t}^m p(\{J_{\ell,t}\})
\end{align} 
and the definition of the Born probabilities. Thus only the mean of $J_{\ell,t}$
depends on the state but not its variance. This is implemented by replacing $J_{\ell,t}\to \langle\hat n_\ell\rangle_t+\tfrac{\xi_{\ell,t}}{2\gamma dt}$ with  $\xi_{\ell,t}$ being Gaussian white noise. Inserting the replacement into the generalized projector in Eq.~(\ref{eq:proj}) yields:
\begin{align}
\hat M(\{J_{\ell,t}\})&=\prod_\ell[\tfrac{2\gamma dt}{\pi}]^{\frac{1}{4}}\exp[-\gamma dt (\langle\hat n_\ell\rangle_t-\hat n_\ell+\tfrac{\xi_{\ell,t}}{2\gamma dt})^2]\\
&=\prod_\ell[\tfrac{2\gamma dt}{\pi}]^{\frac{1}{4}} \exp[-\gamma dt (\langle\hat n_\ell\rangle_t-\hat n_\ell)^2-\xi_{\ell,t}(\langle\hat n_\ell\rangle_t-\hat n_\ell)].\nonumber
\end{align}
We now expand the exponential to first order in $dt$, keeping in mind that $\xi_{\ell,t}^2=\gamma dt$. It yields
\begin{align}
\hat M(\{J_{\ell,t}\})\approx \frac{1}{\mathcal{N}}\left[ 1-\sum_\ell \frac{\gamma dt }{2}(\hat n_\ell - \langle\hat n_\ell\rangle_t)^2-\xi_{\ell,t}(\hat n_\ell - \langle\hat n_\ell\rangle_t) \right],\nonumber
\end{align}
where $\mathcal{N}$ is a constant prefactor. Applying this expansion to the wave function and subsequently normalizing it, shows that $\mathcal{N}$ drops out and one obtains the stochastic Schrödinger equation (SSE)~\cite{alberton2021enttrans,Cao2019,Bernard_2018} (with $\hat m_{\ell,t}=\hat{n}_\ell-\langle \hat{n}_\ell \rangle_t$)
\begin{align} \label{eq:QSD}
    d\ket{\psi_t} &= \Big( -i dt \hat{H} + \sum_\ell  [\xi_{\ell,t} \hat m_{\ell,t}-\tfrac{\gamma dt}{2}\hat m_{\ell,t}^2]  \Big) \ket{\psi_t }.
\end{align}

\subsection{Fermion replica master equation and replica field theory} In order to  average over the random measurement outcomes, we utilize the replica framework~\cite{buchhold2021effective,poboiko,Fava2023} with the generalized projector. We introduce $r=1,...,R$ replicas of the fermion Hilbert space and their fermion operators $ \hat c^{\dagger(r)}_\ell, \hat c^{(r)}_\ell$ and $\hat n_\ell^{(r)}=\hat c^{\dagger(r)}_\ell \hat c^{(r)}_\ell$. The measurement-averaged density matrix for fixed $R,M>0$ is
\begin{align}
    \rho_{R,M}&=\overline{p(\{J_{\ell,t}\})^R \otimes_{r=1}^M |\psi_t\rangle\langle\psi_t|} =\overline{\Tr(|\tilde\psi_t\rangle\langle\tilde\psi_t|)^{R-M}\hspace{-2mm}\otimes_{r=1}^M  |\tilde\psi_t\rangle\langle\tilde\psi_t|}\nonumber\\
    &=
    \Tr_{r>M}\tilde\rho, \text{ with } \ \ \tilde\rho= \overline{\otimes _{r=1}^R|\tilde\psi_t\rangle\langle\tilde\psi_t|}.
\end{align}
The trajectory average amounts to an integration over all possible outcomes $J_{\ell,t}\in \mathds{R}$. It relates the \emph{nonlinear average} of $M$-replicated wave functions, weighted with Born probability $p(\{J_{\ell,t}\})^R$ to a \emph{linear average} of $R$-replicated, unnormalized wave functions. The latter is well-defined for $R\ge M$. The case $R=1$ is obtained via analytic continuation. 

The infinitesimal evolution $\partial_t\tilde\rho$ is obtained by performing the measurement average and expanding the result up to $\mathcal{O}(dt)$. Consider the unnormalized, trajectory averaged density matrix $\tilde\rho= \overline{\otimes _{r=1}^R|\tilde\psi_t\rangle\langle\tilde\psi_t|}$. In an infinitesimal time step $t\to t+dt$, for each replica one unitary operator $\exp{-i\hat{H}dt}$ and one generalized projector $\hat M(\{J_{\ell,t}\})$ from Eq.~(\ref{eq:proj}) act from the left and from the right on the density matrix. We attribute each operator a replica label $r$, indicating the replica Hilbert space, and an additional sign $\sigma=\pm$ indicating whether it acts from the left or from the right onto $\tilde\rho$, i.e., $\hat{c}_\ell\rightarrow \hat{c}_{\ell,\sigma}^{(r)}$. This simplifies the notation: each pair of fermions $\hat{c}_{\ell,\sigma}^{(r)}, \hat{c}_{\ell',\sigma'}^{(r')}$ is independent if any of the labels differ from each other. The evolution is then $\tilde\rho_{t+dt}=U^{\otimes2R}\overline{\hat{M}(\{J_{\ell,t}\})^{\otimes 2R}}\tilde\rho_t$, with $U^{\otimes2R}=\mathds{1}-idt\sum_{r=1}^R (H^{(r)}_+-H^{(r)}_-)$ and the averaged replica projector. Before the average,
\begin{align}
    &\hat{M}(\{J_{\ell,t}\})^{\otimes 2R}=\prod_\ell \left[\frac{2\gamma dt}{\pi}\right]^{\frac{R}{2}}\exp\left[ -\gamma dt\sum_{\sigma=\pm,r}(J_{\ell,t}-\hat{n}_{\ell,\sigma}^{(r)})^2 \right]\nonumber\\
    &=\prod_\ell \left[\frac{2\gamma dt}{\pi}\right]^{\frac{R}{2}} e^{-\gamma dt R(2 \tilde{J}_{\ell,t}^2+L^2)} \exp{ \tfrac{\gamma dt}{2R}\left(\hat N_{\ell,+}+\hat N_{\ell,-}\right)^2}.\label{eq:average_proj}
\end{align}
Here we have introduced $\hat N_{\ell,\sigma}=\sum_{r=1}^R\hat{n}_{\ell,\sigma}^{(r)}$ and $\tilde{J}_{\ell,t}=J_{\ell,t}-\frac{1}{2R}\left(\hat N_{\ell,+}+\hat N_{\ell,-}\right)$  and we have used the fermion property that $(\hat{n}_{\ell,\sigma}^{(r)})^2=\hat{n}_{\ell,\sigma}^{(r)}$ and that $\sum_\ell \hat{n}_{\ell,\sigma}^{(r)}=L^2/2$ is an exact identity for a $U(1)$ symmetric, half-filled lattice.  Taking the trajectory average over all possible measurement outcomes amounts to integrating over all possible values for each $J_{\ell,t}$. Since $J_{\ell,t}$ for different time steps and lattice sites are independent, we can perform this integration for each infinitesimal evolution operator individually. The Born probabilities are implemented at the end by the correct replica limit. Thus, performing the average on $\hat{M}(\{J_{l,t}\})^{\otimes 2R}$ amounts to a Gaussian integral, yielding
\begin{align}
    \overline{\hat{M}(\{J_{l,t}\})^{\otimes 2R}} &= \exp{- \gamma dt RL^2+\frac{\gamma dt}{2R}\left(\hat N_{\ell,+}+\hat N_{\ell,-}\right)^2}\nonumber \\
    &\overset{dt\to0^+}{=}\mathds{1}-\frac{\gamma dt}{R}\sum_\ell\left[R^2-\tfrac{1}{2}\left(\hat N_{\ell,+}+\hat N_{\ell,-}\right)^2\right].
\end{align}
The replica quantum master equation (rQME) is readily obtained acting the operators $\hat N_\ell$ on $\tilde\rho_t$ as indicated by the labels $\sigma=\pm$
\begin{align}\label{eq:replica_QME}
\partial_t\tilde{\rho}=&i\big[\tilde{\rho},\hat{H}_R\big] -\tfrac{\gamma}{R}\sum_{\ell} \left(\{ \hat N_{\ell}(R-\hat N_\ell), \tilde{\rho} \}+\tfrac{1}{2}\big[\hat N_\ell,\big[\hat N_\ell, \tilde{\rho} \big] \big]\right).
\end{align}
Here, $\hat H_R=\sum_{r, \sigma} \hat H^{(r)}_\sigma$ and $\hat N_\ell=\sum_{ \sigma} \hat N_{\ell,\sigma}=\sum_{r, \sigma} \hat n_{\ell,\sigma}^{(r)}$ are the sum over $r$-replicated Hamiltonian and number operators. Both are quadratic in fermions and thus invariant under replica-rotations. The competition between $\hat H_R$ and $\hat N_\ell$ drives the MIPT: the latter prefers a replica-aligned local particle density $\hat N_\ell=0,R$, while the Hamiltonian instead aims to maximize the kinetic energy, i.e., pushes $\hat N_\ell\to \tfrac{R}{2}$.

The fermion rQME can be mapped to a Keldysh path integral~\cite{Sieberer_2016,Kamenev2011}: at each infinitesimal time step $t\to t+dt$ fermion coherent states are inserted on both sides of the density matrix $\tilde\rho_t$. Each fermion operator acting on coherent states at time $t$, is then replaced by Grassmann variables $c_{\ell,\sigma}^{(r)\dagger}, c_{\ell,\sigma}^{(r)}\to \bar\psi_{\ell,t,\sigma}^{(r)}, \psi_{\ell,t,\sigma}^{(r)}$. Here $\sigma$ again denotes whether the operator acts from the left ($\sigma=+$) or from the right ($\sigma=-$) onto the density matrix~\cite{Sieberer_2016}. For the monitoring part of the action, this can be readily inferred from Eq.~\eqref{eq:average_proj}, yielding 
\begin{align}
    iS_{\text{meas}}=\int dt\sum_\ell \left[\sum_{r=1}^R(\bar\psi_{\ell,t,+}^{(r)}\psi_{\ell,t,+}^{(r)}+\bar\psi_{\ell,t,-}^{(r)}\psi_{\ell,t,-}^{(r)})\right]^2.
\end{align}
Now, one performs a canonical rotation into the fermion Keldysh basis via~\cite{Kamenev2011} $
    \psi_{\ell,t,\alpha}^{(r)}=(\psi_{\ell,t,+}^{(r)}-(-1)^\alpha\psi_{\ell,t,-}^{(r)})/\sqrt{2}$ and $\bar\psi_{\ell,t,\alpha}^{(r)}=(\bar\psi_{\ell,t,+}^{(r)}+(-1)^\alpha\bar\psi_{\ell,t,-}^{(r)})/\sqrt{2}$, where $\alpha=1,2$ is the Keldysh index. Introducing the vector $\psi_{t,\ell}=\{\psi_{\alpha,t,\ell}^{(r)}\}$ and similarly $\bar\psi_{t,\ell}$ and drawing the spatial continuum limit $(t,\ell)\to(t,\mathbf{x})=X$ yields the partition sum $Z=\int\mathcal{D}[\{\psi,\bar\psi\}]\text{ exp}(iS_\psi)$ and action
\begin{align}\label{eq:Ferm_Keld}
    S_\psi=\int_{X} \left\{\bar\psi_X G_0^{-1}\psi_X -\tfrac{i\gamma}{2R}\Tr[(\sigma^K_x\psi_X\bar\psi_X)^2]\right\}.
\end{align}

    The free part of the action $\sim \int_X \bar\psi_X G_0^{-1}\psi_X$ with the free Green's function in Keldysh structure $G_0^{-1}$ is familiar from textbooks, see, e.g., Ref.~\cite{Kamenev2011}. Here $\psi_X=\{\psi^{(r)}_{\alpha,X}\}$ is a $2R$-Grassmann vector and the trace runs over replica and Keldysh index; the Pauli matrix $\sigma^K_x$ acts on the Keldysh index $\alpha$. At half-filling, the bare Keldysh Green's function in momentum space $P\equiv (\omega, \mathbf{p})$ is $G_0(P)=\delta_{r,r'}[(\omega-\epsilon_{\mathbf{p}})\mathds{1}-i0^+\sigma_z^K]^{-1}$ with dispersion $\epsilon_{\mathbf{q}}=2\text{cos}(p_x)+2\text{cos}(p_y)$.

The action $S_\psi$ is reminiscent of the Keldysh action for disordered fermions with $U(1)$-symmetry~\cite{Kamenev2011}. The quartic fermion vertex, however, displays two crucial differences compared to static disorder, which are intrinsic to a monitored, i.e., dynamic, theory. Firstly, the structure in Keldysh space is modified by $\sigma_x^K$ (disorder: $\sigma_x^K\to\mathds{1}$). This reduces the symmetry of rotations in Keldysh-replica space: it gaps out $R^2-1$ rotation modes, which can be treated perturbatively. The result is an \emph{emergent space-time invariance} at long wavelengths and a dynamical critical exponent $z=1$. Secondly, the vertex is space-time local: measurements, unlike disorder, vary in \emph{time and space}. This eliminates the possibility of fermions to interfere with their time-reversed partners and gives rise to \emph{two local} long-wavelength modes. 
 
\subsection{Bosonic theory --\\  Saddle point and Gaussian fluctuations}
We proceed by obtaining a bosonic action via a Hubbard-Stratonovich decoupling. We multiply the partition function $Z=\int\mathcal{D}(\{\psi_X\})e^{iS_\psi}$ by the identity
\begin{align}
    1=\int\mathcal{D}(\{Q_X\})\delta(Q_X-2\psi_X\bar\psi_X).
\end{align}
Here $Q_X\in\mathds{C}^{2R\times2R}$ is a $2R\times2R$ Hermitian matrix, while $\psi_X\bar\psi_X$ is a $2R\times 2R$ matrix of Grassmann bilinears and $\delta(...)$ is the Dirac $\delta$-function. Shifting this identity into the Grassmann integration, we use the $\delta$-function to replace the fermion bilinear matrices $\psi_X\bar\psi_X\to Q_X$ in the quartic term. For the fermion bilinears, one has the identity
\begin{align}
 \Tr\left[(\sigma_x\psi_X\bar\psi_X)^2\right]=-\left(\Tr\left[\sigma_x\psi_X\bar\psi_X\right]\right)^2,
\end{align}
which has no counterpart for a general $Q_X$-matrix structure. In terms of $Q_X$, both terms are relevant, i.e., due to the spacetime-local nature of the matrix $Q_X$ both correspond to slow, long-wavelength fluctuations in a Hartree-Fock decomposition of the four fermion vertex. This is in contrast to conventional disordered problems~\cite{Kamenev2011}. We therefore choose a symmetric decoupling of both channels in this work.

We thus have $Z=\int\mathcal{D}[\{\psi_X,Q_X\}]\delta(Q_X-2\psi_X\bar\psi_X)e^{iS_{\psi-Q}}$ with
\begin{align}\label{eq:Ferm_Keld2}
    S_{\psi-Q}=\int_{X} \left\{\bar\psi_X G_0^{-1}\psi_X -\tfrac{i\gamma}{8R}\left(\Tr[(\sigma_xQ_X)^2]-(\Tr[\sigma_xQ_X])^2\right)\right\}.
\end{align}
In order to implement the $\delta$-function, we use its integral representation in terms of a $2R\times2R$ Hermitian matrix $\Xi_X$, 
\begin{align}
    \delta(Q_X-2\psi_X\bar\psi_X)\sim \int\mathcal{D}[\{\Xi_X\}]e^{\frac{i}{2}\Tr[\Xi_X(Q_X-2\psi_X\bar\psi_X)]}.
\end{align}
As common for nonlinear sigma model approaches, we replace the hard constraint resulting from an integration over $\Xi$ by a soft constraint~\cite{Sachdev2011}: we replace $\Xi$ by its saddle point value, obtained by taking the variation with respect to $Q_X$. The remaining action is quadratic in Grassmann variables. Integrating them out and replacing $\Xi$ by the saddle point value yields
\begin{align}\label{eq:bosonaction}
iS_Q=&iS_0+\tfrac{\gamma}{8R}\int_X\Big\{\left[\Tr(\sigma^K_xQ_X)\right]^2-\Tr\left[(\sigma^K_xQ_X)^2\right]
\Big\}
\end{align}
with  $iS_0=\int_X\Tr\ln(G_0^{-1}+i\tfrac{\gamma}{2R}Q_X)$. 
The replica-diagonals $Q_{\alpha\beta, X}^{(rr)}=\psi^{(r)}_{\alpha,X}\bar\psi_{\beta,X}^{(r)}$ represent the physical fermion bilinears, such as, e.g., the fermion density  $n_X^{(r)}=\tfrac{1}{2}(Q_{12, X}^{(rr)}+Q_{21, X}^{(rr)})$ and 
 provide access to the connected correlation function~\cite{buchhold2021effective,poboiko} \begin{align}
C(\mathbf{x}-\mathbf{x}',t)=\langle n_{\mathbf{x},t}^{(r)}[n_{\mathbf{x}',t}^{(r)}-n_{\mathbf{x}',t}^{(r')}]\rangle=\overline{\langle \hat n_{\mathbf{x}}\hat n_{\mathbf{x}'}\rangle-\langle \hat n_{\mathbf{x}}\rangle\langle\hat n_{\mathbf{x}'}\rangle}.
\end{align}
For free fermions it further enables computing an approximate value for the entanglement entropy $S_A$ of a subregion $A$. The latter can be expressed in terms of a series incorporating all even ($2n-$th) order cumulants $C_A^{(2n)}$ of the particle number $\hat N_A=\sum_{\ell\in A}\hat n_\ell$ in subregion $A$~\cite{KlichLevitov, KlichEntanglement}. It yields $S_A=2\zeta(2)\mathcal{C}_A^{(2)}+\mathcal{O}(C_A^{(2n\ge4)})$. In order to obtain an analytical expression for the entanglement entropy and the mutual information, we approximate the series by the first order term, which is proportional to $C_A^{(2)}=\int_{\mathbf{x},\mathbf{x}'\in A} C(\mathbf{x}-\mathbf{x}',t)$. Here the time $t$ is taken at the of the simulation, i.e., at a $(2+1)$-dimensional space time volume with `absorbing' temporal boundaries~\cite{poboiko}.

In order to determine the saddle point of the action $S_Q$ we take the variation with respect to $Q_X$. The locality of $Q_X$ in space-time simplifies the variation compared to disordered fermions and yields the common result~\cite{Kamenev2011} 
\begin{align}
    0\overset{!}{=}\tfrac{\delta S_Q}{\delta Q_X}=\tfrac{i\gamma}{2R}\left[\left(G_0^{-1}+i\tfrac{\gamma}{2R}Q\right)_{X,X}^{-1}-\tfrac{1}{2}\sigma_xQ_X\sigma_x+\tfrac{1}{2}\sigma_x\Tr(\sigma_xQ)\right].\nonumber
\end{align}
The equation is solved by $Q_X=\Lambda=\delta_{r,r'}\sigma_z$, which also implements causality for the saddle point fermion Green's function $G^{-1}_{\text{sp}}\equiv\left(G_0^{-1}+i\tfrac{\gamma}{2R}\Lambda\right)_{X,X}^{-1}\delta_{r,r'}$. It also implements a half-filled Keldysh Green's function and fulfills the nonlinear constraint $Q_X^2=\mathds{1}$. The result amounts to the self-consistent Born approximation~\cite{Kamenev2011} with an elastic mean-free time $\tau_{\text{el}}=R/\gamma$. 

In order to extract density correlation functions we expand the action around the saddle point by setting $Q_X= \Lambda+ \delta Q_X$ and then expanding $S_Q$ to quadratic order in $\delta Q_X$. The zeroth order term is a constant and the first order term vanishes, leaving us with 
\begin{align}
    S_{\delta Q}^{(2)}=\tfrac{\gamma^2}{8R^2}&\int_{X,X'} \Tr(G_{\text{sp}}(X-X')\delta Q_{X'}G_{\text{sp}}(X'-X)\delta Q_{X})\nonumber\\
    &-\tfrac{\gamma}{8R}\int_X\left\{ \Tr\left[(\sigma_x\delta Q_X)^2\right]-\left[\Tr(\sigma_x\delta Q_X)\right]^2\right\}.
\end{align}
This action can be significantly simplified: the Green's functions $G_{\text{sg}}$ are diagonal in replica- and Keldysh index. Furthermore, integrals over $G_R(X)G_R(-X)$ and $G_A(X)G_A(-X)$ are zero due to causality. We thus end up with a separation of the action into replica diagonal modes $\Phi^{r}_X=\delta Q_X^{rr}$ and off-diagonal modes $\delta\tilde Q_X^{r,r'}$. Furthermore, one finds a pairing of $\Phi^{12}$ and $\Phi^{21}$ in Keldysh index. It is convenient to introduce \emph{classical} and \emph{quantum} fields $\Phi^{q,r}_X=\delta Q^{21,rr}_X$ and $\Phi^{c,r}_X=\delta Q^{21,rr}_X$. In Fourier space, we find
\begin{align}
    iS_\Phi=&-\tfrac{1}{8}\int_P (\Phi^{c,r}_P, \Phi^{q,r}_P)\left(\begin{array}{cc} 0& \mathcal{G}^{-1}(P)\\ \mathcal{G}^{-1}(-P)&  0\end{array}\right)\left(\begin{array}{c}\Phi^{c,r}_{-P} \\ \Phi^{q,r}_{-P}\end{array}\right)\nonumber\\
    &+\tfrac{\gamma (1-\delta_{r,r'})}{8R}\int_P (\Phi^{c,r}_P, \Phi^{q,r}_P)\left(\begin{array}{cc}1& 1\\ 1& 1\end{array}\right)\left(\begin{array}{c}\Phi^{c,r'}_{-P} \\ \Phi^{q,r'}_{-P}\end{array}\right).
\end{align}
The inverse propagator here is defined via the integral
\begin{align}
    \mathcal{G}^{-1}(P)=[\tfrac{\gamma^2}{R^2}\int_{P'} g^R(P+P')g^A(P')]-\tfrac{\gamma}{R},
\end{align}
which for $\omega, \mathbf{p}^2\ll\gamma$ and $R\to1$ yields $\mathcal{G}^{-1}(P)\approx -i\omega+\tfrac{2}{\gamma}\mathbf{p}^2$. 
Computing the particle number fluctuations as defined previously thus yields 
\begin{align}\label{eq:Gauss_ent}
     C(P)=\frac{\text{Re}(\mathcal{G}(P))}{2+4\gamma \text{Re}(\mathcal{G}(P))}.
\end{align}
When computing the Fourier transform of this equation, one needs to consider that the observable is measured at the temporal boundary of the $2+1$-dimensional evolution. This amounts to integral boundaries $t\in [0,\infty)$ instead of the usual $t\in (-\infty,\infty)$. Without this boundary, the prefactor for $S_A$ would be $\frac{\pi}{6\gamma}$. The correct expression is a factor of $2$ larger~\cite{poboiko}. Thus, in the Gaussian approximation and for $A=L/2\times L$ we find the cumulant and the leading order approximation for the entanglement entropy to be
\begin{align}\label{eq:Gaussian}
C(P)=\tfrac{\mathbf{p}^2}{\gamma(\omega^2+4\mathbf{p}^2)} \text{ and } S_A=\tfrac{\pi}{3\gamma}\ln(\pi L/2)L.
\end{align} 

\begin{figure}
\includegraphics[width=\linewidth]{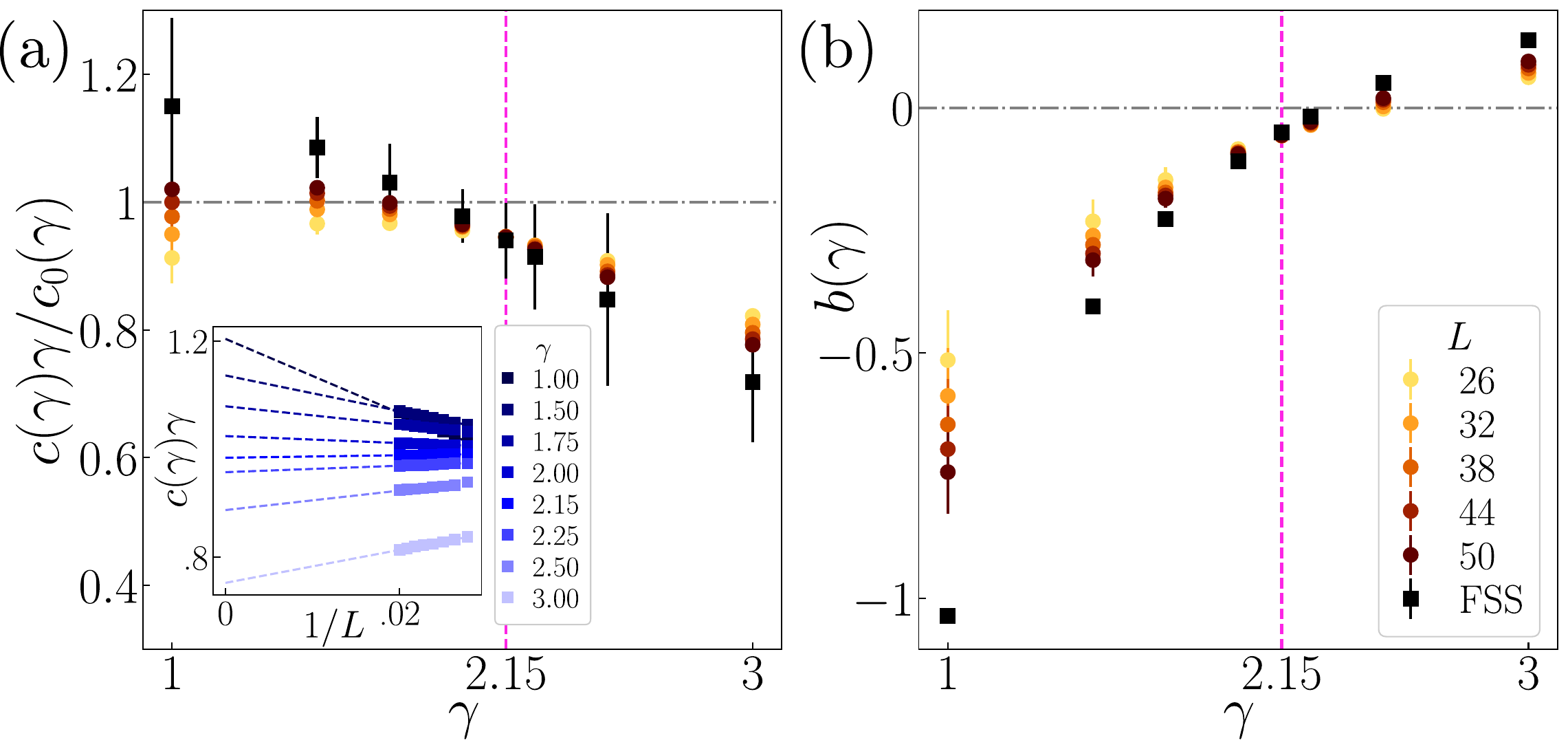}
\caption{\label{fig:cft_params} {\bf Entanglement entropy scaling}. Computing the half-system entanglement entropy $S(A=L\times L/2)$ and fitting it to the form $S(A=L\times L/2)= c(\gamma) L\ln(L)+b(\gamma)L$ predicted by the saddle point analysis yields: (a) A sharp crossing point of the prefactor $c(\gamma)$ at $\gamma=2.15$. The black squares result from an extrapolation of $c(\gamma)$ to $L\to \infty$. Agreement with the result of Eq.~\eqref{eq:Gaussian}, $c_0(\gamma)=\pi/3\gamma$, is found for intermediate $\gamma$ around $\gamma\approx 2.15$. Inset: Finite-size scaling data for $\gamma c(\gamma)$. (b) The residual entropy $b(\gamma)$ displays a sharp crossing point at $\gamma=2.15$ and vanishes in the vicinity of this point. }
\end{figure} 

\subsection{Nonlinear sigma model} The long-wavelength fluctuations of the bosonic matrix $Q_X$ correspond to all symmetry-compatible rotations $Q_X=\mathcal{R}_{X} \Lambda \mathcal{R}_X^{-1}$ of the saddle point $\Lambda$. Each $\mathcal{R}_X$ performs local rotations in $2R\times 2R$-dimensional replica and Keldysh space compatible with a U$(2R)$-symmetry. Since $\Lambda\sim \sigma_z$ in Keldysh space, $2R^2$ generators are trivial, i.e., those $\sim \mathds{1}, \sigma_z$ in Keldysh space. The resulting rotations act in U$(2R)/$U$(R)\times $U$(R)$ symmetric space and are parametrized by generators $\sigma_x\Theta_x, \sigma_y\Theta_y$, where $\Theta_{x,y}$ are Hermitian $R\times R$ matrices acting on replica coordinates. We set $\mathcal{R}_X=\mathcal{R}_{x,X}\mathcal{R}_{y,X}$ with $\mathcal{R}_{\eta,X}=\exp(\tfrac{i}{2}\sigma_\eta\Theta_\eta)$.

Inserting this parameterization into the measurement part and using identities for Pauli matrices to compute the trace in Keldysh indices, we find
\begin{align}
   \Tr[(\sigma_xQ_X)^2]&=-2\Tr_R\cos(2\Theta_{y,X})\approx 4\Tr_R(\Theta_{y,X}^2),\nonumber\\
(\Tr[\sigma_xQ_X])^2&=(2\Tr_R[\sin(\Theta_{y,X})])^2\approx 4 (\Tr_R\Theta_{y,X})^2.
\end{align}
Here $\Tr_R$ denotes the trace over the remaining replica indices. We have expanded the nonlinear functions since the action does not vanish for the traceless part of $\Theta_{y,X}$, i.e., for $\tilde\Theta_{y,X}=\Theta_{y,X}-\frac{1}{R}\Tr_R\Theta_{y,X}$, which thus remains gapped independently of the value of $R$. The remaining U$(1)$ trace degree of freedom $\theta_{y,X}\equiv \Tr_R\Theta_{y,X}$ has a gap that vanishes in the limit $R\to1$ and is thus not integrated out. It does not enter the computation of the entanglement entropy and remains diffusive in the limit $R\to1$, where it describes the average diffusion of particles.

The logarithm in the action $S_Q$ is rewritten~\cite{Kamenev2011}
\begin{align}
    \Tr\left[\ln(G_0^{-1}+i\tfrac{\gamma}{2R}Q_X)\right]&=\Tr\left[\ln(\mathcal{R}_X^{-1}G_0^{-1}\mathcal{R}_X+i\tfrac{\gamma}{2R} \Lambda)\right]\nonumber\\
    &=\Tr\left[\ln(\mathds{1}+G_{\text{sp}}\mathcal{R}_X^{-1}[G_0^{-1},\mathcal{R}_X])\right],
\end{align}
with $G_{\text{sp}}=\left(G_0^{-1}+i\tfrac{\gamma}{2R}\Lambda\right)^{-1}$. The logarithm is expanded to leading order in derivatives. The temporal derivative yields
\begin{align}
\Tr\left[G\mathcal{R}_X^{-1}i\partial_t\mathcal{R}_X \right] &=-\tfrac12\Tr\left[\sigma_x^K\sin(\Theta_{y,X})\mathcal{R}_{x,X}^{-1}\partial_t\mathcal{R}_{x,X} \right]\\
&\approx-\tfrac12\Tr_R\left[\Theta_{y,X} e^{-\frac{i}{2}\Theta_{x,X}}(\partial_te^{i\Theta_{x,X}})e^{-\frac{i}{2}\Theta_{x,X}} \right].\nonumber
\end{align}
The spatial derivative appears only in quadratic order due to inversion symmetry
\begin{align}
    \Tr((\nabla Q_X)^2)=2\Tr_R[(\nabla  e^{i\Theta_{x,X}})(\nabla e^{-i\Theta_{x,X}})].
\end{align}
Integrating out the gapped relative replica fluctuations $\tilde \Theta_{y,X}$ in the Gaussian approximation and rescaling time by a factor $\sqrt{2}$ yields
\begin{align}\label{eq:NONLS}
    iS_U=-\frac{g}{2}\int_X [\partial_t U\partial_t U^{-1}+\nabla U\nabla U^{-1}]
\end{align}
with $U\in\text{SU}(R)$ and $g=\sqrt{2}/(8\gamma)$. 
 It also emerges in common disorder problems for the chiral unitary class AIII~\cite{EversMirlin}. Setting $R\to1$ and in $2+1$ dimensions, the one-loop renormalization group (RG) flow for the dimensionless coupling $\tilde g=g l$ is (RG scale $l$)~\cite{WegnerRG,HikamiRG,gade1991the}
\begin{align}\label{eq:RGscale}
    \partial \tilde g(l)/\partial\ln(l)=\tilde g(l)-\tfrac{1}{4\pi}\ \Rightarrow \ \tilde g(l)=\tfrac{1}{4\pi}+(g-\tfrac{1}{4\pi})\tfrac{l}{l_0}.
\end{align}
Here, $l_0=1$ is the UV-scale (dimensionless lattice spacing) and $\tilde g(l_0)=g$. The flow equation predicts a MIPT at a \emph{critical monitoring rate} $\gamma_{c,\text{th}}=\tfrac{\pi}{\sqrt{2}}\approx 2.22$. For  $\gamma<\gamma_{c,\text{th}}$ ($\gamma>\gamma_{c,\text{th}}$), the theory flows to weak (strong) coupling.

\subsection{Effective Non-Hermitian Hubbard model}
While the continuum action in the boson framework in Eq.~\eqref{eq:bosonaction} emphasizes the link to Anderson physics, it obscures a higher SU$(2R)$ that is present whenever fermions hop on a bipartite lattice, composed of sublattices $A$ and $B$, such as, e.g., the square lattice. To reveal it, we utilize the third quantization procedure~\cite{ProsenThird,Essler2016} and reexpress the rQME by viewing $\tilde\rho\to||\tilde\rho\rangle\rangle$ as a state vector in Hilbert space with evolution $\partial_t ||\tilde\rho\rangle\rangle=i\hat{\mathcal H}||\tilde\rho\rangle\rangle$. Introducing an index $\sigma$ denoting operators acting on the left ($\sigma=+1$) or on the right ($\sigma=-1$) of $\tilde{\rho}$ and performing a unitary transformation $U_A=\exp(-i\pi \sum_{\ell \in A}\hat N_\ell^-)$ that acts only on the $(\sigma=-1)$-degrees on sublattice $A$, one finds (half-filling)
\begin{align}\label{eq:diss_Hub}
    \hat{\mathcal H} = \sum_{r,\sigma}\hat{H}^{(r,\sigma)} + \frac{i\gamma}{2R}\sum_\ell \Big[\Big(\sum_{r,\sigma}\hat n_\ell^{(r,\sigma)}\Big)-R\Big]^2-i\gamma L^2 R .
\end{align}
The \emph{Hamiltonian} $\hat{\mathcal H}$ is invariant under rotation of $2R$ flavors $(r,\sigma)$ and a global phase and thus displays a U$(1)\times$ SU$(2R)$ rotation symmetry. It resembles an attractive Hubbard model with $2R$ flavors, where the attractive coherent interaction vertex is replaced by a dissipative one through the evolution operator $\exp(i\hat{\mathcal H} t)$ acting on $||\tilde\rho\rangle\rangle$. The stationary state $||\tilde\rho\rangle\rangle_{t\to\infty}$ corresponds to precisely the eigenstate of $\hat{\mathcal H}$ whose eigenvalue features the smallest imaginary part. For the case $R=1$, the stationary state and the excitation spectrum of the dissipative Hubbard model in Eq.~\eqref{eq:diss_Hub} have been explored in the context of particle decoherence~\cite{Essler2016}. While decoherence leads to a trivial stationary state, any $R>1$ enforces a relaxation towards a nontrivial stationary state. We note that $||\tilde\rho\rangle$ is not a conventional wave function. For instance, its norm (the trace) is computed with respect to a 'flat' state $\langle 1||$~\cite{ProsenThird} and it is only preserved in the limit $R=1$. 

A full SU$(2R)$-symmetric Keldysh action is obtained by either computing a 'standard', single-contour path integral for the generator $\exp(i\hat{\mathcal{H}}t)$ or by performing the sublattice transformation $U_A$ on the Keldysh action in Eq.~\eqref{eq:replica_QME} before taking the spatial continuum limit. In the latter case, an additional particle-hole transformation on the $\sigma=-$ contour with $\psi\to\bar\psi, \bar\psi\to -\psi$ yields the Keldysh field theory with a full SU$(2R)$ symmetry~\footnote{This symmetry was also recently discovered for for monitored fermion models with $N_F$ fermion flavors~\cite{Fava2024}.}. This formulation of the problem in terms of a dissipative Hubbard model or SU$(2R)$-symmetric Keldysh theory opens a previously unanticipated direction on monitored fermions. It remains a challenging task to explore the analytical predictions from this approach and its consequences on the MIPT of monitored fermions. Below, we demonstrate by numerical simulations that indeed the picture of the fermion MIPT may need to be refined.

\section{Results}
\subsection{Numerical implementation} The
SSE \eqref{eq:QSD} is quadratic in $\hat{c}_\ell,\hat{c}_\ell^\dagger$ and efficiently simulated with Gaussian states~\cite{Cao2019,alberton2021enttrans}
\begin{equation}\label{eq:gstate}
    \ket{\psi_t}=\prod_{1\le s\le L^2/2} c^\dagger_s\ket{0}, \quad c^\dagger_s=\sum_{1\le\ell\le L^2} \psi^s_{\ell,t} c^\dagger_\ell.
\end{equation}
 The single-particle wave function $\psi^s_{\ell,t}\in\mathds{C}$ of fermion $s$ at site $\ell$ implicitly depends on the history of noise events $\{\xi_{\ell,t'<t}\}$. Normalization of the wave functions implies $\sum_\ell (\psi_{\ell,t}^s)^*\psi_{\ell,t}^{s'}=\delta_{s,s'}$, i.e., $\psi_{t}^\dagger \psi_{t}^{\phantom{\dagger}}=\mathds{1}$.
 The trajectory evolution thus amounts to updating $\psi_{t}$, which is done by a Suzuki-Trotter decomposition of Eq. (\ref{eq:QSD}) ~\cite{Cao2019,alberton2021enttrans}. Up to first order in $dt$ and up to an overall normalization, this leads to the matrix update
\begin{equation}\nonumber
    \psi_{t+dt}=\text{diag}\left( e^{\xi_{1,t}+\frac{\gamma dt}{2} (2\langle \hat{n}_1\rangle_t-1)},\dots,e^{\xi_{N,t}+\frac{\gamma dt}{2} (2\langle \hat{n}_N\rangle_t-1)} \right)e^{-ihdt}\psi_{t}.
\end{equation}
Here, $h_{\ell,\ell'}=-\delta_{\ell\text{ n.n.}\ell'}$ is the nearest-neighbor hopping matrix. 
Normalization is obtained by performing a QR-, i.e., Gram-Schmidt, decomposition $\psi=QR$ and redefining $\psi\equiv Q$. 
We initialize each $\ket{\psi_{t=0}}$ in a random Gaussian state at half-filling and evolve it until our observables, i.e., the von Neumann entanglement entropy $S(A)$ 
 and the statistics of $\psi_{\ell,t}^s$, reach stationary values.
 
For subsystems $A,B$ we compute the trajectory averages 
\begin{align}
     S(A)= -\Tr (\overline{\rho_A \ln \rho_A}), \ 
     \mathcal{I}(A,B)= S(A)+S(B)-S(A\cup B).\nonumber
     \end{align}
The reduced density matrix $\rho_A=\Tr_{\bar A}|\psi_t\rangle\langle\psi_t|$ is obtained by tracing over the complement of $A$. Here, we take each subsystem as a strip of size $A=L\times l_A$, see Fig.~\ref{fig:overview}b. The wave functions $\psi_\ell^{s}$ are characterized by their inverse participation ratio (IPR) $P_q$, its variance $\sigma_q$ and anomalous dimension $D_q$,
\begin{equation}\label{eq:IPR}
    P_q = 2 L^{-2}\sum_{s,\ell} |\psi^s_{\ell,t} |^{2q}\sim L^{-D_q(q-1)},  \ \sigma_q^2=\text{var}(\ln P_q).
\end{equation}

\begin{figure}
\includegraphics[width=\linewidth]{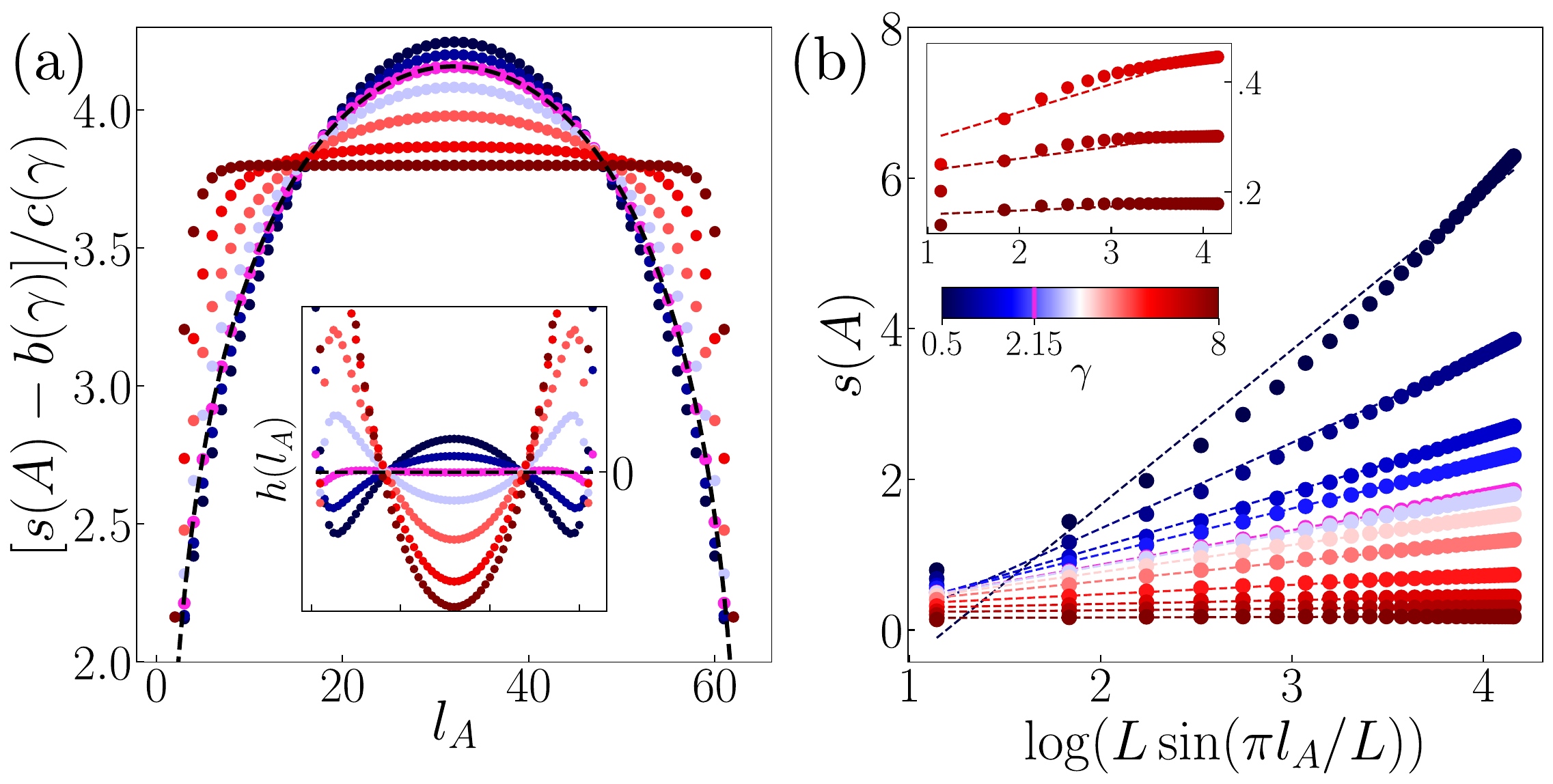}
\caption{\label{fig:cft} {\bf Emergent conformal symmetry}. (a) Entropy line density for strips $A=L\times l_A$ compared with Eq.~\eqref{eq:cft}, for $L=64$. Deviations are visible for all curves except for $\gamma=2.15$. Inset: Deviation from Eq.~\eqref{eq:cft}. Here $h(l_A)=[s(A)-b(\gamma)]/c(\gamma)-\ln[L\sin(\pi l_A / L)]$ is zero only when Eq.~\eqref{eq:cft} holds. (b) Entropy line density $s(A)$ as a function of $\log[L\sin(\pi l_A/L)]$ (circles) for $L=64$ and different measurement strengths. Dashed lines represent the linear fit results. A clear linear behaviour is seen only for $\gamma\approx2.15$. Inset: Same plot for large measurement strengths $\gamma\ge4$, highlighting the deviations from Eq.~\eqref{eq:cft}.}
\end{figure}

\subsection{Entanglement and mututal information} The entanglement and mutual information in this work are exclusively computed from the von Neumann entanglement entropy. For a Gaussian state, it is computed from the correlation matrix $\langle \psi_t|c^
\dagger_{\ell}c^{\phantom{\dagger}}_m|\psi_t\rangle=(\psi_t^{\phantom{\dagger}}\psi_t^\dagger)_{\ell,m}$.  In order to compute the von Neumann entanglement of a subsystem $A$, one considers the correlation matrix $(\psi_t^{\phantom{\dagger}}\psi_t^\dagger)|_A$ restricted to subsystem $A$ and computes its eigenvalues $\{\lambda_\alpha\}|_A$. The von Neumann entanglement entropy for the subsystem is then
\begin{equation}\nonumber
S(A)=-\sum_\alpha\lambda_\alpha\log\lambda_\alpha+(1-\lambda_\alpha)\log(1-\lambda_\alpha).
\end{equation}

Numerical simulations underpin two entanglement regimes: for weak measurements the entanglement entropy grows as $S(A=L\times L/2)= c(\gamma) L\ln(L)+b(\gamma)L$, which is reminiscent of a 2D metallic state and consistent with the field theory result in Eq.~\eqref{eq:Gaussian}. For strong measurements an area law $S(A)= b(\gamma) L$, i.e., $c(\gamma)=0$, is observed in Fig.~\ref{fig:overview}d.
We extract the prefactor of the sub-extensive growth in three ways: (i) For any given $L$, we perform a linear fit of $S(A)/\tilde L$ vs $\ln \tilde L$ for all even $\tilde L$ values in $\tilde L\in[2,L]$ and with $L$ values up to $L=50$. (ii) We use the relation $c(\gamma)\approx S(A)/(L\ln L)$ for each $\gamma$ value and $L$ values up to $L=64$. (iii) Using the data of the entanglement entropy $s(A)$ as a function of the subsystem size $A=L\times l_A$, we perform a linear fit of $s(A)$ vs $\log[L\sin(\pi l_A/L)]$ for different $\gamma$ and $L$ values. All methods are in close agreement, see Appendix~\ref{sec:comp_a} for a comparison. For intermediate values of $\gamma$, the numerical prefactor is in good agreement with the Gaussian approximation $c_0(\gamma)= \tfrac{\pi}{3\gamma}$ in the metallic phase, see Fig.~\ref{fig:cft_params}a\footnote{Approximating the entanglement entropy with the leading order cumulant $C^(2)_A$ has been shown to become unreliable for large and small $\gamma$ in one spatial dimension~\cite{poboiko}.}. The residual entropy $b(\gamma)$ in this regime remains small ( $|b(\gamma)|\le10^{-1}$) and crosses zero around $\gamma\approx2.15$, see Fig.~\ref{fig:cft_params}b. We have furthermore extrapolated the infinite-system values of $c(\gamma)$ and $b(\gamma)$ (as computed with method (i)) by means of a finite-size scaling analysis. In particular, we have performed a linear fit $c(\gamma)$ = $m/L+c(L\to\infty)$ and thus extracted $c(L\to\infty)$ and similarly for $b(L\to\infty)$, see inset of Fig.~\ref{fig:cft_params}a. Finite size scaling of the metallic prefactor $c(\gamma)$ reveals a sharp crossing point at $\gamma\approx 2.15$ in Fig.~\ref{fig:cft_params}a. The existence of a crossing point at non-zero $c(\gamma)$ is unanticipated in the Anderson theory~\cite{poboiko,poboiko2023measurementinduced}, which predicts a critical point with area law entanglement.

We confirm this feature by extracting the entropy line density $s(A)=S(A)/L$ for fixed $L=64$ and variable strip size $A=L\times l_A$ in Fig.~\ref{fig:cft}. We compare it with the formula
\begin{equation}\label{eq:cft}
 s(A) = c(\gamma)\ln\left[\tfrac{L}{\pi}\sin\left(\tfrac{\pi l_A}{L}\right)\right]+b(\gamma)
\end{equation}
for a quasi-one-dimensional geometry with conformal symmetry~\cite{Calabrese_2004,Casini_2009}. At $\gamma\approx2.15$, we find a perfect match with Eq.~\eqref{eq:cft}, while the entanglement curve is flatter (sharper) for $\gamma>2.15$ ($\gamma<2.15$). This emergent symmetry at $\gamma\approx2.15$ is a major result of this work and it is uncommon for Anderson-type transitions.

\begin{figure}
\includegraphics[width=0.7\linewidth]{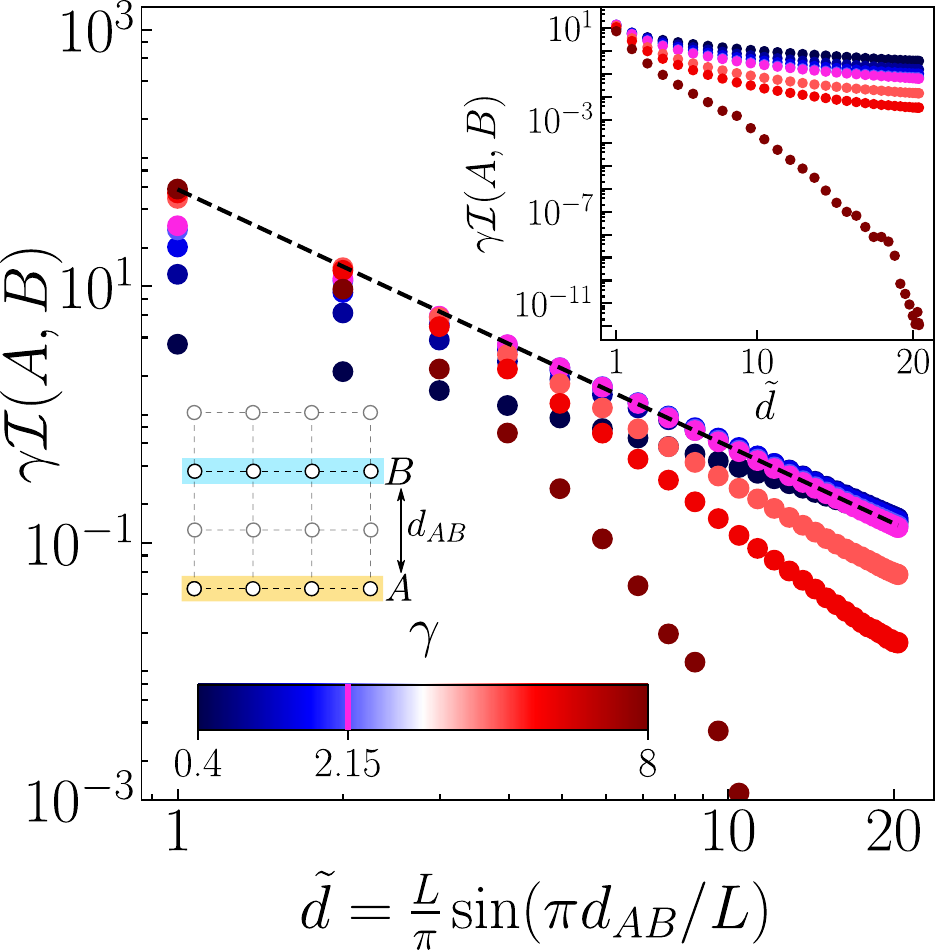}
\caption{\label{fig:MI} {\bf Mutual Information}. A scaling collapse of the mutual information as a function of the chord distance $\tilde{d}$ confirms the saddle point prediction $\mathcal{I}(A,B)\sim  \tilde{d}^{-2}/\gamma$ from Eq.~\eqref{eq:Gaussian} in the metallic phase for $\gamma\le2.15$. Data is computed for a fixed system size of $L=64$. Above this value deviations towards localization are visible. Inset: Unrescaled data on a semi-log scale, highlighting the exponential decay of the mutual information for $\gamma\gg2.15$.}
\end{figure}

A further measure for the MIPT is the mutual information $\mathcal{I}(A,B)$, which provides an upper bound for the correlations between disjoint subsystems $A, B$~\cite{wolf2008area}. For free fermions it is determined by particle number fluctuations between $A$ and $B$~\cite{KlichLevitov,KlichEntanglement}. To leading order, $\mathcal{I}(A,B)=4\zeta(2)\int_{\mathbf{x}\in A, \mathbf{x}'\in B}C(\mathbf{x}-\mathbf{x}',t)$. For two strips of size $L\times1$, separated by a distance $d_{AB}$, see Fig.~\ref{fig:MI}, Eq.~\eqref{eq:Gaussian} predicts $\mathcal{I}(A,B)\sim  d_{AB}^{-2}/\gamma$. This is confirmed by a scaling collapse of $\mathcal{I}(A,B)$ for $\gamma\le2.15$ in Fig.~\ref{fig:MI}. For $\gamma>2.15$, one instead observes deviations from the predicted results and an evolution towards an exponential decay $\log (\mathcal{I}(A,B))\sim -d_{AB}$, confirming localization  \cite{poboiko2023measurementinduced}. 

\subsection{Multifractality} A striking feature of localization transitions is the emergence of multifractality~\cite{EversMirlin}. The multifractal exponent $D_q$ defined in Eq.~\eqref{eq:IPR} characterizes strong fluctuations of single-particle wave functions. For wave functions in $D$ dimensions, it  distinguishes a metallic phase, with $D_q=D$, from a localized phase where $D_q\to 0$, both independent of $q$. At a multifractal critical point $0<D_q<2$ becomes a non-trivial function of $q$.

\begin{figure}
\includegraphics[width=\linewidth]{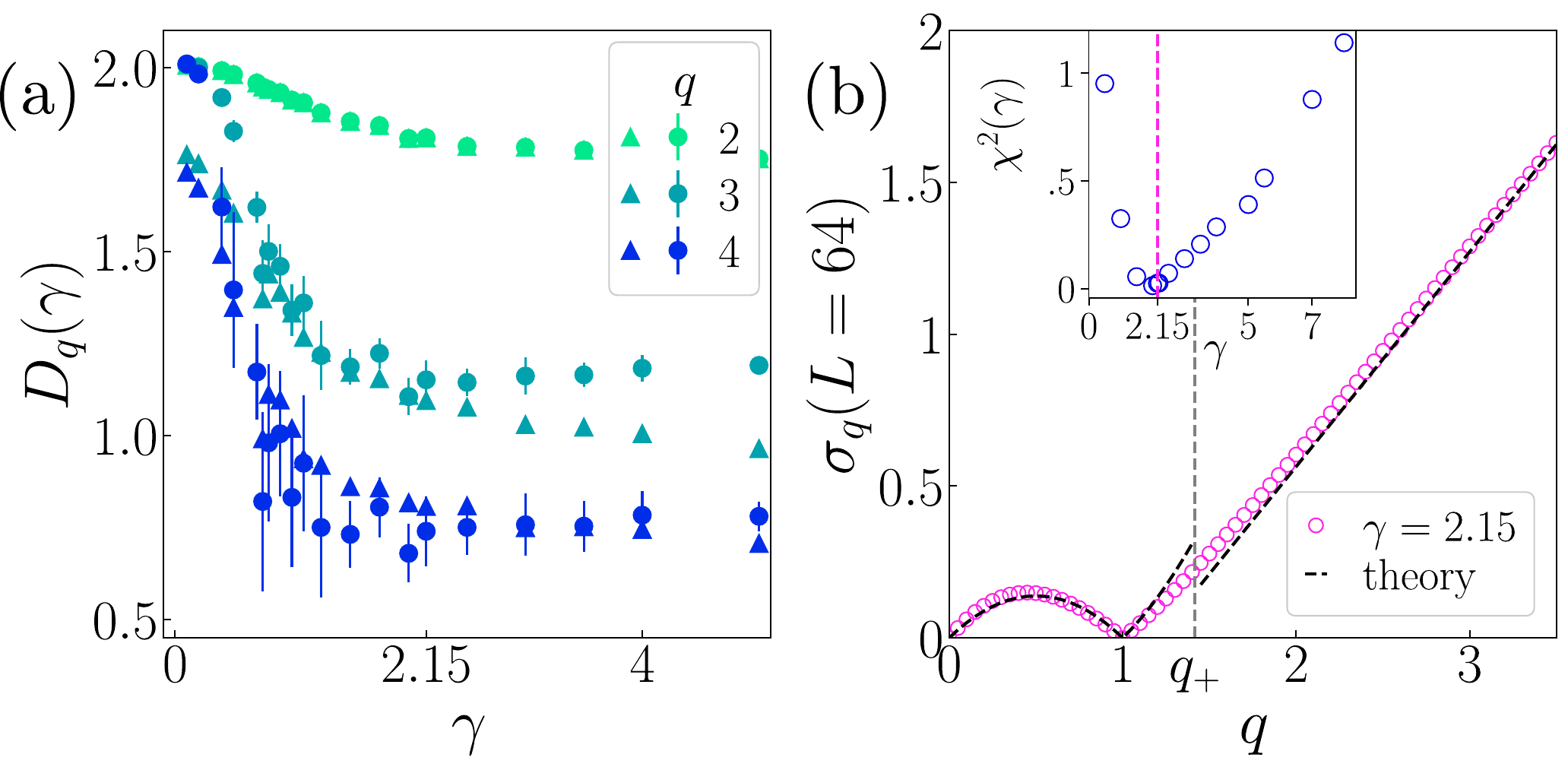}
\caption{\label{fig:mf} {\bf Multifractality}. (a) Multifractal exponent $D_q$ for different $\gamma, q$, estimated by evaluating Eq.~\eqref{eq:IPR} at $L=64$ (triangles) and from fit of $\text{ln}(P_q)/(1-q)$ versus $\text{ln} (L)$ for even $L\in[10,64]$ (circles). (b) The variance $\sigma_q$ of the distribution $\mathcal{P}(\ln P_q)$ both for the monitored fermions (pink circles) and the analytical prediction from the three-dimensional Anderson model~\cite{Mildenberger} (grey dashed lines). Inset: The $\chi^2(\gamma)$-deviation between theory and simulation, see Eq.~\eqref{eq:chi}, is minimal and close to zero for $\gamma\approx2.15$.}
\end{figure}

We extract $D_q$ in two different ways, (i) from a fit of $\text{ln}(P_q)/(1-q)$ versus $\text{ln} (L)$ for even $L\in[10,64]$ and (ii) evaluating Eq.~\eqref{eq:IPR} at $L=64$. We find qualitative agreement
between both methods. In Fig.~\ref{fig:mf}a we show $D_q$ for a range of $q$ and $\gamma$ values. It has sigmoid behaviour: it saturates to the metallic value $D_q\to2$ for $\gamma\to0$ for all $q$ and slowly decays with increasing monitoring strength, revealing an extended region of multifractal behaviour. We note that $D_q$ approaches the value $D_q\to0$ only very slowly, i.e., for both $L$ and $\gamma$ large, pointing towards a large localization length.

A second characteristic of multifractality is the variance $\sigma_q$ of $ \ln P_q$.  For symmetry class AI in $D=3$ analytics predict
\begin{equation}\label{eq:IPRvar}
    \sigma_q^{\text{Th}}=\left\{\begin{array}{ll}
    2\pi b q|(q-1)|& \text{ for } |q|<q_+ \\
     q/q_+& \text{ for } |q|>q_+
    \end{array}\right., 
\end{equation}
with 
$b\approx0.088$ and $q_+\approx \sqrt{2}$~\cite{Mildenberger}. Indeed, as pointed out in Ref.~\cite{poboiko}, the one-loop flow equations for the chiral symmetry class AIII in the replica limit $R\to1$ resemble those for the orthogonal class AI in the replica limit $R\to0$~\cite{WegnerRG}. Therefore we compare our data and Eq.~\eqref{eq:IPRvar}in Fig.~\ref{fig:mf}b. The quality of agreement is quantified through a $\chi^2$ value
\begin{equation}\label{eq:chi}
    \chi^2(\gamma)=\sum_i |\sigma_{q_i}(\gamma)-\sigma_{q_i}^{\text{Th}}|^2/\sigma_{q_i}^{\text{Th}}
\end{equation}
for a grid of $q_i\in(0,1)$. Similarly, we have estimated the slope of the linear behaviour of our data and checked the closeness to the predicted value of $1/q_+$. Both quantities predict a best match with the theory at the symmetry point $\gamma\approx 2.15$~\footnote{We note that while the multifractal exponents plateau around $\gamma\approx2.15$, the variance $\sigma_q(\gamma)$ still is sensitive to changes in the measurement strength.}.

\subsection{Purification}
\begin{figure}
\includegraphics[width=\linewidth]{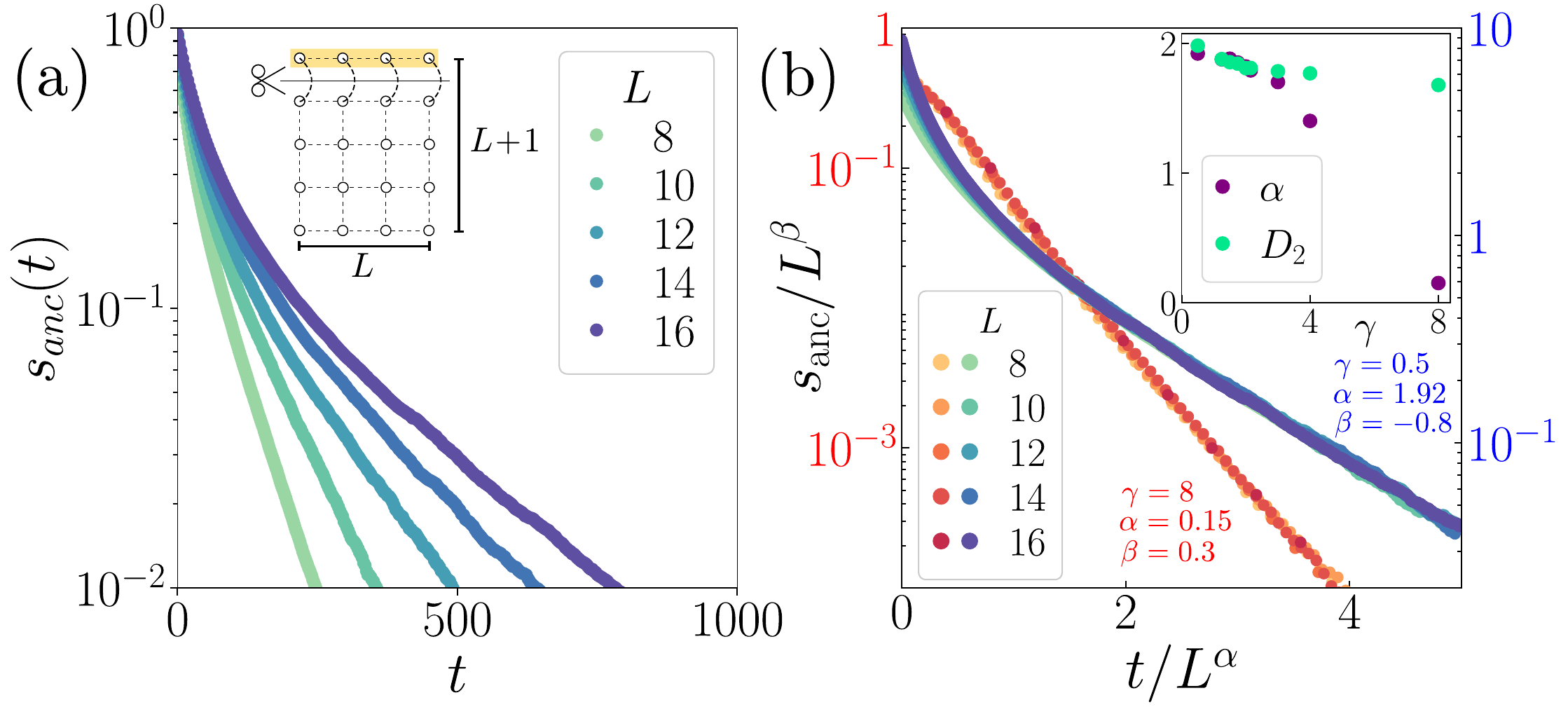}
\caption{\label{fig:purification} {\bf Purification time scale}. (a) Unrescaled purification dynamics of an $L\times1$ ancilla for $\gamma=0.5$ and different $L$ values. The drawing represents the system+ancilla setup. (b) Finite size scaling of the purification rate $\tau_{\text{pur}}\sim L^{\alpha}$ for strong (red) and weak (blue) monitoring. Inset: Comparison of the purification rate exponent $\alpha$ and the multifractal exponent $D_2$ for different $\gamma$.}
\end{figure}

MIPTs can be revealed from the purification of a mixed initial state~\cite{Gullans2020,gullans2019,GullansPurNonHerm, loio2023}. We extract the purification time scale $\tau_{\text{pur}}\sim L^\alpha$ by initially entangling the $L\times L$ lattice with an $L\times 1$ ancilla~\cite{gullans2019scalable}, i.e., by bringing fermions into a superposition between ancilla and system, see Fig.~\ref{fig:purification}. For times $t>0$ only the $L\times L$ lattice undergoes monitoring described by the SSE~\eqref{eq:QSD}. The purity is provided by the entropy density of the ancilla $s_{\text{anc}}\equiv S(L\times 1)/L\sim \text{exp}(-t/\tau_{\text{pur}})$. 
While the system is entangled with the ancilla and thus itself in a mixed state, the global state of the system \emph{and} ancilla is in a Gaussian pure state. The global system can thus be numerically simulated as outlined above. The purification time scales of the system, however, are typically much longer than the time needed for other observables to reach stationary values. The purification simulations are thus computationally more costly and only smaller system sizes up to linear dimension $L=16$ are performed. 
To determine the scaling of the purification rate $\tau_{\text{pur}}$ with the system size $L$, we performed a finite-size scaling analysis. This was done by rescaling time as $t\to t/L^\alpha$ and the ancilla entanglement density as $s_{\text{anc}}\to s_{\text{anc}}/L^\beta$. Then, we found the $\alpha$ and $\beta$ values which would collapse the curves for different $L$ values, as shown in Fig.~\ref{fig:purification}b.

The obtained purification exponent $\alpha$ closely resembles the multifractal exponent $D_2$ for $\gamma\lesssim2.15$, while $\alpha\to0$ rapidly for $\gamma\gtrsim2.15$. This behavior can be rationalized:  purifying a particle (or hole) shared by system and ancilla amounts to localizing it by measurement. The probability of finding the particle in one measurement step depends on how its wave function is scrambled, i.e., on its multifractal structure. 

\section{Outlook} Monitored free fermions in 2D experience a localization-delocalization MIPT. On a bipartite lattice, the microscopic $R$-replica model can be mapped to both an SU$(R)$-symmetric nonlinear sigma model in 2 + 1 dimensions and an SU$(2R)$-symmetric non-Hermitian Hubbard model in two dimensions. Both models offer different perspectives and introduce an unanticipated twist in our understanding of MIPTs of fermions, which we explore through numerical simulations. While wave function multifractality, typical for Anderson transitions, matches NL$\sigma$M predictions at $\gamma=2.15$, this point also exhibits a higher symmetry consistent with conformal invariance -- unexpected in Anderson theory. This calls for a closer inspection of the link between monitored fermions and equilibrium localization transitions, and whether monitored fermions may represent a new nonequilibrium universality class. Further, we note that the mapping to the SU$(2R)$-Hubbard model is exclusive to continuous monitoring, raising the question of whether projective and continuous measurements correspond to different universality classes -- a simultaneous work~\cite{poboiko2023measurementinduced} reports agreement with NL$\sigma$M predictions for projective measurements in 2D. Clarifying the similarities and differences between projective and continuous monitoring may be an interesting future direction~\cite{Romito2020,Schomerus2019}.

Probing multifractality via purification makes 2D monitored fermions an ideal test bed to explore such questions. A future goal is to devise experimental schemes to explore the MIPT~\cite{Monroe2021} through appropriate feedback protocols~\cite{AbsorbingCircuitHuse,IadecolaControl,Buchhold2022a,Piroli2023, sierant2023controlling}.

\begin{acknowledgments}
  We thank S. Diehl, I.~V. Gornyi, C.~M. Jian, K. Klocke, I. Poboiko, M. Szyniszewski, X. Turkeshi and J.~H. Wilson for fruitful discussions.
  We acknowledge support from the Deutsche Forschungsgemeinschaft (DFG, German Research Foundation) under Germany's Excellence Strategy Cluster of Excellence Matter and Light for Quantum Computing (ML4Q) EXC 2004/1 390534769, and by the DFG Collaborative Research Center (CRC) 183 Project No. 277101999 - project B02. The code for our numerical computations was implemented in Julia~\cite{bezanson17}. 

  \emph{Note added} -- During the completion of the manuscript, we learned about related works on monitored fermions in $D\ge2$ that appeared simultaneously~\cite{poboiko2023measurementinduced, jin2023measurementinduced}.
\end{acknowledgments}

\bibliography{EntEnt}

\begin{appendix}
    \section{Comparison of entanglement scaling}\label{sec:comp_a}
\begin{figure}
\includegraphics[width=0.8\linewidth]{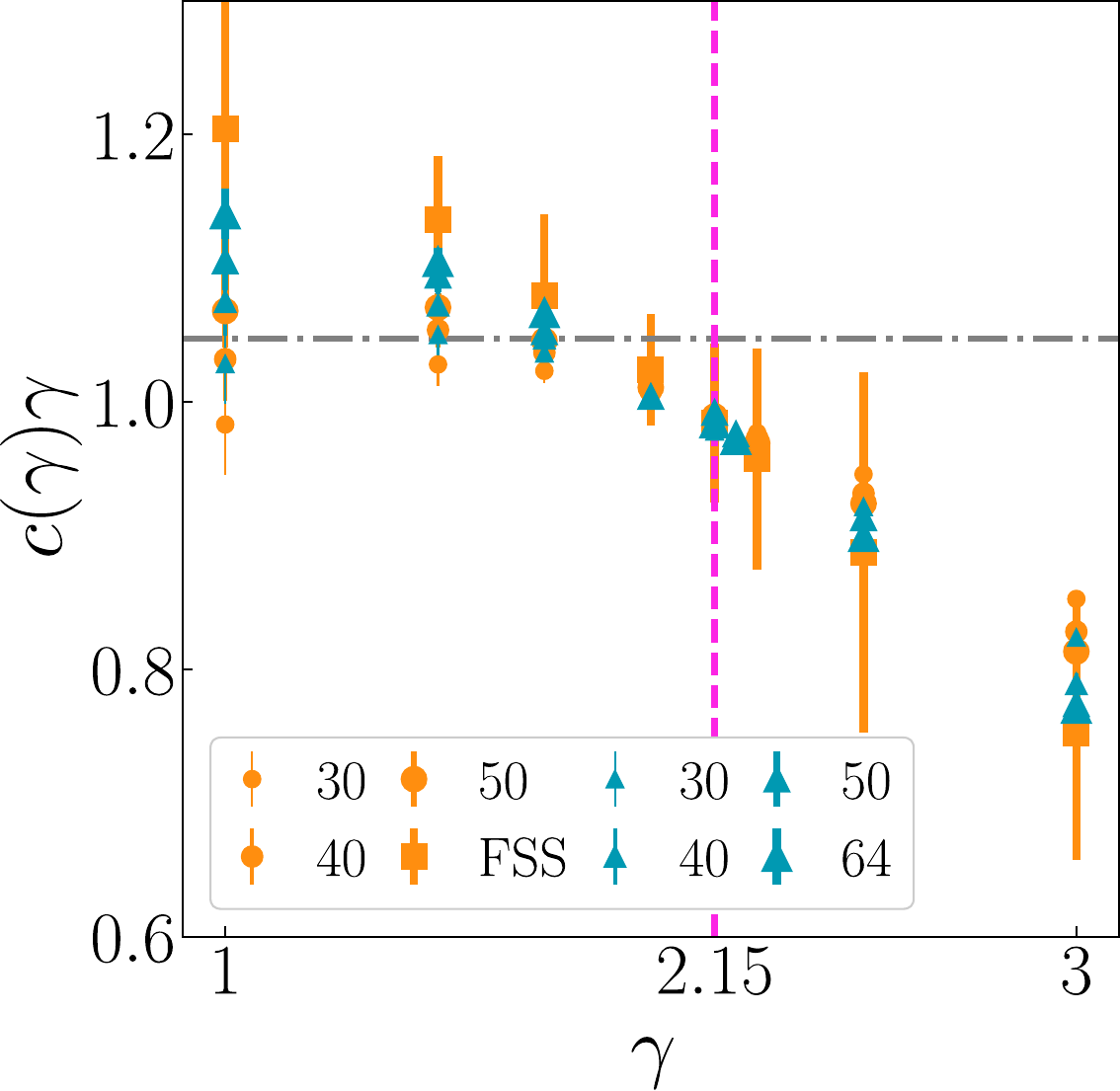}
\caption{\label{fig:app_cft_c} {\bf Comparison of entanglement entropy scaling}. Prefactor of the logarithmic scaling of the half-system entanglement entropy estimated from the entanglement entropy data as a function of system size $L$ (orange circles, equivalent to Fig.~\ref{fig:cft_params} main text) and subsystem size $l_A$ (blue triangles). Despite using different methods, both data  agree well and in particular predict the same sharp crossing point at $\gamma\approx2.15$.}
\end{figure}

We present here a comparison between different procedures to extract the logarithmic scaling prefactor $c(\gamma)$ of the half-system entanglement entropy $S(A = L\times L/2)$. The value of $c(\gamma)$ in Fig.~\ref{fig:cft_params} is computed by means of a linear fit of the entanglement entropy as a function of system size $L$. As outlined in the main text, an alternative estimate of $c(\gamma)$ is obtained by using the data of the entanglement entropy as a function of the subsystem size $A=L\times l_A$, i.e. via a linear fit of $s(A)=S(A)/L$ vs $\log[L\sin(\pi l_A/L)]$ for $L=64$ and $l_A\in [1, 32]$. Although both methods are clearly different, they yield the same sharp crossing point at $\gamma\approx2.15$ and are in close agreement between each other, see Fig.~\ref{fig:app_cft_c}. 

\end{appendix}

\end{document}